# An ECR hydrogen negative ion source at CEA/Saclay: preliminary results.

R. Gobin*, P-Y. Beauvais, O. Delferrière, R. Ferdinand, F. Harrault, J-M. Lagniel.
Commissariat à l'Energie Atomique, CEA-Saclay, DSM/DAPNIA/SEA
91191 Gif sur Yvette Cedex, France
*e-mail: rjgobin@cea.fr

*Abstract:*

The development of a high intensity negative ion source is part of a considerably larger activity presently undergoing at CEA Saclay in the field of high intensity linear accelerators. Preliminary studies toward the construction of a 2.45 GHz ECR H- ion source have been performed for few months. This new test bench takes advantage of our experience on the French high intensity proton source SILHI. In the new source, the high-energy electrons created in the ECR zone are trapped by a dipole magnetic filter. A rectangular 200 mm long plasma chamber and an intermediate iron shield are used to minimize the magnetic field in the extraction region. A second magnetic filter separates electrons and negative ions in a 10 kV extraction system. To reduce the electron/H- ratio, the plasma electrode is slightly polarized. The design allows future evolutions such as cesium injection, higher energy extraction and plasma diagnostics. The installation of the source is now in progress. The first helium plasma has been produced for few weeks to verify the electron separator behavior. The design, computations and the first results of the source are presented.

## I - Introduction

Potential applications of high current accelerators include the production of high flux neutron beams for spallation reactions (ESS), future reactors, nuclear waste treatment, exotic ion facilities or neutrino and muon production for high-energy particle physics. The high intensity beams for these accelerators may reach an energy as high as 1 GeV. In France, CEA and CNRS have undertaken an important R&D program on very high beam power (MW class) light-ion accelerators for several years. Part of the R&D efforts are concentrated on the IPHI (High Intensity Proton Injector) [1] demonstrator project. This 10 MeV prototype of linac front end will accelerate CW beam currents up to 100 mA. It will consist of an intense ion source, a radio frequency quadrupole (RFQ) and a drift tube linac (DTL). The High Intensity Light Ion Source (SILHI) development, based on the 2.45 GHz ECR plasma production, has been performed for several years leading to a great experience in high current proton beam production. Taking into account this advantage, CEA which is involved in the ESS studies, decided to develop a hydrogen negative ion source also based on the ECR plasma production.

Section II gives an overview of the SILHI proton source performance in CW and pulsed mode. The new hydrogen negative ion source is described in section III which also includes magnetic and trajectory computations. Electron separator efficiency and proton density have been measured and preliminary results are reported in section IV. Then the conclusion presents, in section V, the future experiments planned to improve the negative ion source performance.

## II - SILHI, High Intensity Proton Source

To summarize the high intensity proton source efficiency, the most significant results are reported hereinafter. The SILHI proton source [2] has been designed to reach a long lifetime and a very high reliability. It operates at 2.45 GHz. The magnetic field $B_{ECR} = 875$ Gauss is produced by 2 coils tunable independently. The quartz RF window has been installed behind a water-cooled bend to escape the beam of electrons produced and accelerated back to the plasma chamber, in the HV extraction system. The RF window works well since the production of the first beam in July 1996. Nevertheless the boron nitride (BN) disc located at the RF entrance is affected by the backstreaming electrons and must be systematically replaced. Its lifetime is estimated to be higher than 1000 hours for ~ 100 mA CW beams, then more than 40 days of continuous operation for such beams.

The plasma is easily obtained when the RF power is larger than 350 W with the standard magnetic field and operating gas pressure ($10^{-3}$ Torr in the plasma chamber). The source is generally operated 5 days a week for 8 hours daily runs. Less than 10 min. are needed each morning to restart the source with a 100 mA CW beam at 95 keV. The tune up time is reduced to 2 min. after a shut down using an automatic procedure. Less than 6 hours are usually needed to obtain the nominal beam parameters after an operation in the source or in the low energy beam transport (LEBT). This recovery time for pumping, HV column conditioning and tune up is mainly induced by the BN disc outgassing under plasma warming.

The best performances are clearly obtained when two ECR zones are simultaneously located at both plasma chamber extremities. The source efficiency increases to 0.145 mA/W (250 mA/cm$^2$) for 850 W RF forward power in these conditions instead of 0.105 mA/W with a single ECR zone at the RF entrance.

Three long runs have been performed to analyze the reliability – availability of the source. In October 1999, with a 75 mA – 95 keV continuous beam, the reliability reached 99.96 % for a 104 H long operation. Only one beam trip occurs during the test. The beam stopped during 2'30" just one hour after the beginning of the statistics leading to a 103 H uninterrupted operation.

In pulsed mode, rather short plasma rise and fall times have been achieved during some preliminary experiments done using a modulation of the 2.45 GHz magnetron power supply. Plasma pulses with 10 µs rise time and 40 µs fall time have been observed.

For a 80 mA CW proton beam, the nominal r,r' rms normalized emittance is lower than 0.3 pi mm mrad and the proton fraction better than 85 % (12 % for $H_2^+$ and 3 % for $H_3^+$). Several measurements have shown strong improvements of the emittance (0.11 pi mm mrad) when a buffer gas ($H_2$, $N_2$, Ar or Kr) is injected in the LEBT. The space-charge compensation (SCC) factor has been measured at several points along the LEBT. A strong dependence on the number of free electrons in the LEBT line was found. The SCC can be much lower than expected without an increased electron production induced by adding heavy gases for example. Low SSC leads to a strong emittance increase in the LEBT.

## III - Negative hydrogen ion source design

Taking into account this experiment on high intensity beams, it has been recently decided to study a new source for negative hydrogen ion production. The aim is to obtain a long lifetime source with high reliability. As demonstrated with SILHI, these conditions could be reached with sources in which the plasma is generated by ECR. In classical sources, filament or antenna lifetime reduces considerably the reliability.

To design the source, several contacts have been undertaken with different national and foreign laboratories (Ecole Polytechnique, CEA Cadarache, CEA Grenoble, CERN, Frankfurt University) which are involved in H- studies. A step by step work has been decided before the final design.

First, the production of hydrogen negative ions has to be demonstrated in volume production mode before to discuss Cesium or Xenon injection or other improvement like Tantalum surface [3].

Extracted negative ions and electrons have to be separated by means of a magnetic dipole. To avoid $H_2$ excited molecule destruction, high energy electrons have to be eliminated in the plasma close to the extraction area by using a magnetic filter. To do that preliminary magnetic calculations have been performed to design the C shape magnetic electron separator and magnetic filter. Otherwise, the axial magnetic field provided by the two coils to reach $B_{ECR}$ has also been calculated as well as the iron shielding. All computations including particle extraction (Fig. 1) have been done with Axcel [4], Opera 2D and 3D codes [5].

This source also operates at 2.45 GHz ($B_{ECR}$ = 875 Gauss) with a water cooled copper plasma chamber. The rectangular (standard WR 284 waveguide) plasma chamber length has been chosen at 210 mm instead 100 mm for SILHI to reach an axial magnetic field as low as possible close to the extraction area in order to limit the amount of high energy electrons in this zone and to insert the C shape electromagnetic filter.

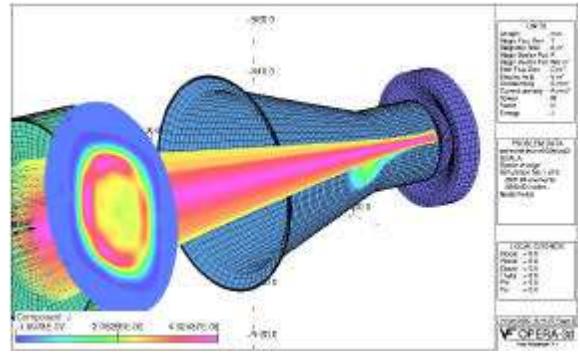

Fig 1: Hydrogen negative ions and electrons separation
The electrons are collected on the electrode

The RF signal is produced by a 1.2 kW magnetron source and is fed to the source via standard rectangular waveguides and an automatic tuning unit. A three section ridged waveguide transition is placed just before the aluminum nitride (AlN) window. This window is located at the RF entrance in plasma chamber. It is protected from backstreaming ions or electrons by a 2 mm BN sheet. The Mo plasma electrode is biased to a few volt power supply.

Figure 2 shows a cross-sectional view of the experimental setup. Several ports have been managed in the plasma chamber for future plasma diagnostics (Langmuir probe, Laser detachment, …).

The source and its ancillaries (power supplies, RF generator, gas injection, …) are grounded and the 10 kV extraction system is installed inside the vacuum chamber. The collector is also linked to an independent HV power supply. The 80 mm aperture C shape tunable electron separator is located inside the vacuum vessel. By using positive or negative HV power supplies, negative and positive extracted beams could be respectively observed.

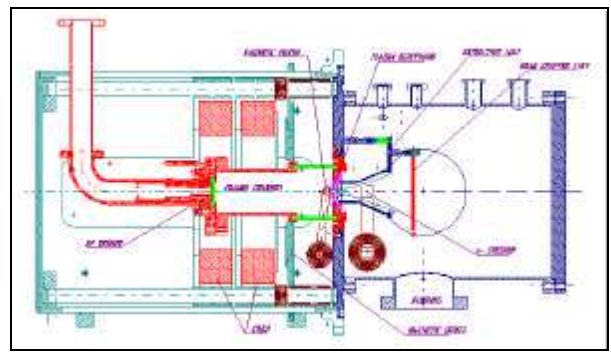

Fig 2: Cross-sectional view of the source
and extraction system

## IV- Preliminary results

To verify the source behavior, the first plasma has been produced by injecting hydrogen gas in the plasma chamber. After some outgassing troubles, it has been quite easily obtained when the ECR zone was located at the RF entrance in the plasma chamber. The RF forward power was 500 W with a reflected power lower than 10 % and with an operating pressure of $3 \cdot 10^{-3}$ Torr in the plasma chamber. As the magnetron pulsation is not yet available, it has been decided to pulse the extraction voltage (8 ms/s) with a continuous plasma for the experiments described hereinafter.

The electron separator efficiency has been verified with an Helium plasma. The total extracted negative charges (20 mA trough a 5 mm diameter plasma electrode) are collected on the extraction electrode for the nominal value of the magnetic dipole.

In Hydrogen, the positive charge density has been also checked with a - 6 kV pulsed extraction voltage. A 10 mA/cm$^2$ beam density has been obtained with the magnetic filter switched off. The density has been reduced by a factor 2 with a Bdl ? 500 Gauss.cm magnetic filter. With a pulsed positive extraction voltage (+ 5kV), the electron density also decreases when the magnetic filter is switched on. It also decreases by tuning the plasma electrode voltage.

## V- Conclusion

Figure 3 shows a general view of the source test stand. The first plasma has been produced for few weeks and any negative hydrogen ion beam has been observed since then. Copper (from plasma chamber body) and carbon pollution due to O-ring attack by the RF has been seen, it is probably the main reason of the hydrogen negative ion destruction. Improvements to avoid this RF attack are in progress. However, the electron separator and the magnetic filter behavior seems quite satisfactory.

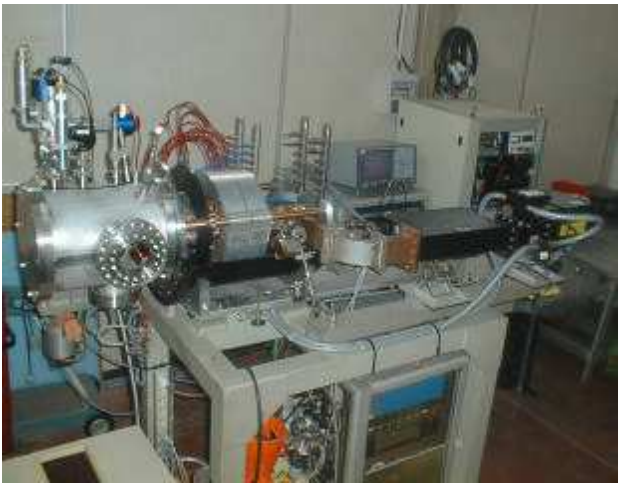

Fig 3: General view of the source test stand

Compare to the SILHI results, the positive ion density is very low because of the magnetic configuration. High negative ion beam intensity could not be produced in such conditions. In the near future, after resolving the O-ring problem, different studies will be done to enhance the plasma confinement in the extraction zone. Computations have been already undertaken by using a multipole magnetic configuration.

For higher energy extraction and to enhance the AlN window lifetime, RF magnetron source pulsation work will be also performed.

When the first step (volume production mode of hydrogen negative ions) will be reached, the second step will consist to improve the performance by Xenon and Cesium injection, Tantalum surface. Then to characterize the beam, the source will be installed close to the SILHI HV platform. The beam will be accelerated through a dedicated accelerator column and analyzed in a diagnostic box located on the platform.

The source efficiency will be analyzed at higher RF power. It is also planned to characterize it at higher RF frequency (10 GHz for example).

## Acknowledgments

Many thanks to the members of the IPHI team for their contributions, especially to G. Charruau and Y. Gauthier for their technical assistance. The authors would also thank M. Bacal, J. Faure, A. Girard, C. Jacquot, G. Melin, J Sherman, K Volk and the CERN source team for their fruitful collaboration and valuable discussions.